# Influence of the temporal contrast on the characteristics of laser-produced ultrafast Kα Hard X-ray source

V. Tcheremiskine*[1,2], Y. Azamoum[1]

[1] LP3 Laboratory, Aix-Marseille University – CNRS, Campus de Luminy, Case 917, 13288 Marseille, France ; [2] P.N.Lebedev Physical Institute of Russian Academy of Sciences, Leninsky prospect 53, Moscow, 119991 Russia

**ABSTRACT**

Characteristics of the intense ultrafast quazi-monochromatic X-ray source at 17.4 keV generated by a multi-TW femtosecond Ti:Sa-laser beam ($\lambda$=800 nm) tightly focused on a thick molybdenum target are investigated within a wide range of laser pulse peak intensities ($10^{16} - 10^{19}$ W/cm$^2$) and temporal contrast ($10^6 - 10^{10}$). Spatial and absolute spectrally-resolved energetic parameters of produced pulsed X ray emission are measured. Fluxes of Mo Kα line radiation reaching $2\times10^9$ photons/sr per laser shot and energy conversion efficiency as high as $3.2\times10^{-5}$ sr$^{-1}$ are obtained at high laser contrast of $>10^9$. The interplay of resonance absorption, vacuum heating and ponderomotive (J×B) heating mechanisms of the laser-plasma interaction is clearly demonstrated by different Kα conversion efficiency dependencies at different laser pulse intensity and contrast parameters. The results are confirmed by examining dependencies on the laser incidence angle. In the relativistic intensity regime, the Kα efficiency shows similar values within the whole examined range of the laser temporal contrast, which varies over 4 orders of magnitude resulting in the electron density scale length $L$ of the pre-formed plasma changing from $L<<\lambda$ to $L>\lambda$. It confirms the dominant role of the J×B mechanism accompanied by the effect of plasma interface steepening under laser radiation pressure in the relativistic interaction regime.

## 1. INTRODUCTION

Recent rapid progress in the high-intensity femtosecond laser technology has boosted investigations in the field of intense laser-matter interactions and resulted in a rapid development of novel coherent and incoherent sources of X-ray emission based on various physical principles. Such sources give the access to a number of unprecedented physical parameters, such as attosecond pulse duration and ultra-high brightness of generated X-ray photon fluxes. [1-4] A great interest to further developments in the field is explained by a number of important and highly promising applications such as, for example, ultrafast diffraction studies and phase-contrast bio-medical imaging. [5,6].

A relatively simple and efficient method to produce intense ultrashort X-ray pulses is realized by delivering of a P-polarised femtosecond laser pulse with intensity of $\geq10^{16}$ W/cm$^2$ to a solid, liquid or cluster target. [7-9] In this case, the dominant fraction of laser radiation is absorbed via collisionless processes, in which suprathermal (hot) electrons are generated with energies exceeding the threshold of K-shell ionization of target material, and even reaching MeV energy level. [10-13] These electrons hit atoms and ions of the target material producing characteristic line and broadband Bremsstrahlung X-ray emission. The Kα-line dominates in the spectrum because of the highest transition probability. Note that energy of emitted Kα-photons can be "tuned" by changing the target material. The X-ray emission is isotropic (at least for laser intensities lower than the ultra-relativistic level). Due to the fact that "pumping" hot electrons are generated only during the intense fs laser pulse, and that afterglow energy relaxation is ultrafast (occurring on the ps time-scale and shorter), the produced X ray bursts have ultrashort picosecond duration and are emitted by a spot, which is comparable in size with the laser focus waist. In principle, the X-ray pulse duration can be reduced down to few hundreds of femtoseconds for proper design and experimental conditions. [14,15] Whereas the source size can be held within the limit of few times the laser focal spot size even for the relativistic interaction regime. [16]

*tcheremiskine@lp3.univ-mrs.fr

The number of Kα X ray photons produced by such source is determined by the amount of generated hot electrons reaching the target with energies exceeding the ionization threshold for the K-shell electrons of target material, as well as by their energy distribution. The latter parameters are determined by the governing mechanism of collisionless absorption of laser radiation by the plasma, which is created at target surface by various kinds of pre-pulses or by rising edge of the main femtosecond pulse before its peak arrives to target. Thus, investigating certain characteristics of the source, for example, the conversion efficiency of the laser energy to the Kα line emission and its dependence on main influencing parameters, one can reveal the kind of dominant laser-target interaction process. As shown in our experiments, the main parameters governing the interaction regime are the laser pulse peak intensity and its temporal contrast. The dependence of Kα conversion efficiency on the angle of incidence of laser radiation to target also brings an important insight to the physics of interaction processes.

The paper is structured as follows. Firstly, we introduce the design of the laser-driven Mo Kα line source of X-ray emission at 17.4 keV developed at the LP3 Laboratory, as well as its spatial and absolute spectral and energetic characteristics. After that, we introduce and discuss the observed dependencies of Kα conversion efficiency on the both key parameters, which determine and change the governing interaction mechanism. We attribute different pieces of these dependences to the action of one of main well-known interaction mechanisms. Then we validate our choice by investigating the efficiency dependence on the angle of incidence of laser radiation to target.

Among the main physical mechanisms governing the laser-plasma interaction we consider the resonance absorption, the vacuum heating (Brunel effect), and the ponderomotive (J×B) heating. [10-13,17] These mechanisms are briefly considered below in the discussion section. The experimental dependencies observed in our work clearly demonstrate the interplay of these main interaction mechanisms with changes in the laser pulse peak intensity and/or contrast. The obtained results are unambiguously understandable taking into account the effect of steepening of the density profile of pre-formed plasma under the radiation pressure of a laser pulse. This effect is clearly seen at laser intensities approaching the relativistic regime.

## 2. EXPERIMENT

The experiments are performed at LP3 Laboratory employing the Ti:Sa laser system "ASUR" (designed by Amplitude Technologies). This system is capable of producing trains of stable (nominally ~1% RMS by energy) high-intensity (~10-TW) pulses of ~28 fs duration, which are generated by two separate amplification lines operating at the pulse repetition rate of 10 Hz and 100 Hz, coordinately. Being focused, the laser beams of the both lines show almost equivalent focal spot size and shape. These beams also exhibit similar temporal contrast parameters. However, all experiments undertaken within the scope of the present paper are performed employing the 10-Hz laser beam, while the 100-Hz laser line is intended mostly for applications.

A high temporal contrast of up to $10^{10}$ with respect to the intensity of nanosecond Amplified Spontaneous Emission (ASE) pedestal is provided with the help of three saturable absorbers (SA), which are incorporated into different stages of the laser amplification chain. [18,19]. These SA are made of thin optical plates (1-2 mm thickness) fabricated from RG-850 Schott color glass. The nominal contrast of ~$3\times10^8$ with respect to ASE is established in basic configuration of the laser system, which includes a single SA incorporated between the stretcher and booster amplifier. In the case that no any SA is used, we also reduce in twice the pump power of the booster amplifier. It results in a significant decrease of the pulse energy seeded into the regenerative amplifier and in the two orders of magnitude lower contrast of output high-intensity fs pulses with respect to that provided by nominal configuration.

Fig.1 introduces the third-order autocorrelation traces registered within a time window of 500 ps for laser pulses generated at different contrast configurations. The temporal contrast of various ultrashort pre-pulses emerging on the ps time scale is measured to be >$10^5$.

As shown in separate measurements performed using various diagnostics of the pulse temporal characteristics (SPIDER, WIZZLER, and a single-shot autocorrelation), the incorporation of SA into the laser amplification chain does not result in considerable changes in the pulse spectrum and duration. [19] Note that price to pay for the high contrast appears to be quite reasonable: ~30% of loss in the pulse energy and by ~3 times degradation of the pulse energy stability with respect to the nominal configuration.

It should be mentioned that significant fluctuations in the laser Contrast Ratio (CR), ranging from $2\times10^{-8}$ to $6\times10^{-8}$ for the nominal configuration (1 SA), and analogous fluctuations for low and high contrast configurations (0 and 3 SA, respectively), are observed in different experimental sessions, which are carried out on different dates (often separated by weeks and months). Such fluctuations are simply explained by small deviations in the laser alignment, ambient temperature, etc. from date to date. Nevertheless, all results presented below are measured at least in two separate experimental sessions and fully agree with each other within the limits of experimental error. Because of above cited fluctuations in the pulse contrast, further in the text we generally assume the following values: $\sim10^6$ for the low contrast configuration (0 SA), from $10^8$ up to $10^9$ for the nominal (1 SA), and from $10^9$ up to $10^{10}$ for the high contrast configuration (3 SA). A full autocorrelation trace of the ASE pedestal, generated in the amplification chain of a very similar to ours laser system, is presented in Ref. [18]. Analogously to it, we assume that the ASE pedestal starts at 5 ns before the main fs pulse with linear growth in intensity, which proceeds during ~3 ns, and is followed by a plateau of ~2 ns duration.

Temporal contrast of various ultrafast pre-pulses (of fs and/or ps duration) emerging on the ns time scale is measured with the help of a fast photodiode and an oscilloscope, which provide a signal-pass bandwidth of 3.5 GHz. In effect, these prepulses are different amplified replicas of the main fs pulse circulating in the laser master oscillator and the regenerative amplifier. At high contrast configuration (3 SA), these measurements show temporal contrast of $\geq10^7$ for such pre-pulses on the ns time scale (with respect to the energy of main fs laser pulse) and confirm the position of starting point of the ASE build up at 5 ns before the main pulse. [19]

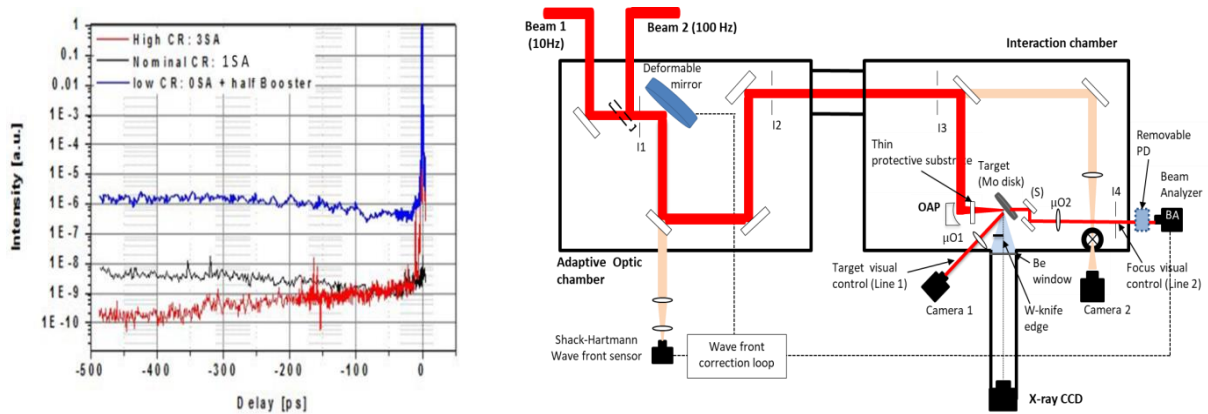

Fig. 1. Third-order autocorrelation traces of the fs pulses generated at different contrast configurations.
Fig. 2. Schematic diagram of the experimental arrangement (to the right).

Fig.2 illustrates the experimental arrangement. Trains of P-polarised laser pulses of ~28 fs duration and of up to ~10 TW peak power are delivered under nominal incidence angle of 45° to the target. The target is a planar thick disk in molybdenum, which rotates around its axis providing each laser impact with a fresh surface. The laser beam is tightly focused on target using an Ag-coated f/3.6 off-axis parabola and an adaptive deformable mirror. Laser pulse peak intensities exceeding $10^{19}$ W/cm$^2$ are obtained at the maximum of a Gaussian-shape focal spot with 10μm waist, which contains ~60% of the total beam energy (about ~40% is within the FWHM diameter of 6 μm). The interaction chamber is evacuated to the pressure of ~$10^{-4}$ mbar. More details on mechanical and optical arrangement of the experiment can be found elsewhere. [20]

Measurements of the parameters of X-ray source emission are provided mainly with the help of a direct detection X-ray CCD camera (Princeton Instruments PIXIS-XB-1024BR). It is mounted at the distance of 92 cm from the source inside an evacuated metal tube connected to the interaction chamber.

The direct detection XCCD cameras allow reconstruction of the spectra of X radiation by determining total charge in a trace of an individual X-ray photon absorption event, which is recorded by the image sensor in the single-hit-per-pixel operation mode. This technique is largely applied to reconstruct spectra within the Soft X-ray range. [21] However, the numerical treatment of images exposed to the more energetic Hard X-ray photons is complicated by the fact that such photon hits simultaneously multiple pixels, and its trace exhibits an arbitrary irregular shape. In our case, for 13×13 μm$^2$ pixel size, three or more pixels are typically hit by a single 17.4 keV photon. To overcome this difficulty, an original home-made numerical code is developed, which is based on a novel treatment algorithm. Note that in practice, the inherent properties of the camera detector result in a rather poor spectral resolution, which is equal in our case to ~0.5 keV for the X-ray photons with energy of 17.4 keV. Nevertheless, as compared with other reported methods, our approach provides more accurate determination of the total charge generated in the image sensor by a separate Hard X-ray photon and allows more accurate spectrally-resolved absolute energetic measurements.

Line emission at 14.4 keV of radioactive Co-57 isotopes is used to provide absolute spectral calibration of our X-ray detection system. The measured value of detection quantum efficiency of 10.2% demonstrates a remarkable agreement with data given by the camera manufacturer.

Reconstructed spectra of the Hard X-ray emission produced by the laser-driven source are introduced in Fig.5. They are in full agreement with those observed by a complete X-ray spectrometer with CdTe detector (Amptek 123-CdTe). However, it should be pointed out that real spectral width of the Kα line emission from a laser irradiated target is limited by several electron volts. Note that more detail consideration of aspects related to the reconstruction of X-ray spectra in our experiments makes the scope of a separate paper.

## 3. RESULTS AND DISCUSSION

An example of the reconstructed spectrum of the Hard X-ray emission produced by the laser-driven Mo Kα line source is demonstrated in Fig.3. This spectrum is obtained in result of treatment of multiple images each exposed to the X-ray photons generated by a single laser impact at energy of 200 mJ on target. To establish the single-hit operation regime, the produced X-ray flux is attenuated by an absorptive filter. This filter consists of Al and Ti plates of 1 mm and 125 μm in thickness, respectively. It is set close to the beryllium window (800 μm thickness) of the interaction chamber. All photons that contribute to the spectrum within the limits of the spectral line registered near 17.4 keV are attributed to the molybdenum Kα emission produced by the source. Energetic conversion efficiency of the laser radiation into the Mo Kα emission is determined by the ratio of the total energy of a burst of emitted Kα photons to the energy of laser impact on target.

The highest value of Kα conversion efficiency reaching $2\times10^{-4}$ into 2π steradian is obtained at laser pulse peak intensity of $\sim5\times10^{18}$ W/cm$^2$ and high temporal contrast of $>10^9$ (configuration with 3 SA). While the highest number of emitted Mo-Kα photons of $1.2\times10^{10}$ (into 2π sr, per shot) is registered at nominal laser contrast of $\sim10^8$ corresponding to the configuration with 1 SA. The reason is that conversion efficiency at relativistic intensities is similar for the both contrast values, but the maximum attainable laser energy on target is by ~30% higher at the nominal contrast. However, as it is demonstrated below, the impact of highly energetic laser pulse with the contrast lower than $10^9$ leads to dramatic expansion of the X-ray emitting spot size, what is extremely undesirable in most of applications.

It should be pointed out that variation of magnitude of the observed ultrashort prepulses (by changing delays of the Pockels cells) within the reasonable limits do not show any considerable effect on the parameters of Kα X-ray emission in our experiments. On contrary, the dependencies of the Kα conversion efficiency on the laser pulse peak intensity observed at different temporal contrast with respect to the intensity of ASE pedestal clearly demonstrate its crucial role. These dependencies are presented in Fig. 4.

Three separate intervals for the laser field intensity are selected on the horizontal axis of the plot in Fig.4. These intervals mark different interaction regimes and correspond to different values of the normalised field amplitude, $a_0 = eE_0/(m_e \omega_{las} c)$, where $E_0$ is the electric field amplitude, $\omega_{las}$ is its oscillation frequency, $e$ and $m_e$ are the charge and the rest mass of electron, respectively, and $c$ is the speed of light. For the Ti:Sa laser radiation of $\lambda$= 800 nm, the normalized field amplitude equals unity (i. e. $a_0$=1) at the laser intensity of $2.2\times10^{18}$ W/cm$^2$. In this case, the maximum velocity of electron oscillations in the laser field (quiver velocity) becomes comparable with the speed of light. Accordingly, we mark in Fig.4 the intensity interval corresponding to the relativistic regime ($a_0 >1$) - by the red colour band, to the sub-relativistic regime ($a_0 \leq 1$) - by light-brown, and to the non-relativistic interaction regime, where $a_0 \ll 1$, by the light-blue band.

As it is seen in Fig.4, the observed values of the Kα conversion efficiency tend to reach a plateau at relativistic laser intensities, where $a_0>1$. It agrees well with results reported by other investigators, who observed the saturation and even decrease of the Mo Kα conversion efficiency at laser intensities beyond $10^{18}$ W/cm$^2$. [22,23] Such effect is explained by the counteracting influence of two physical processes. On one hand, with the increase of laser intensity at relativistic regimes, the efficiency of Kα photon production experiences a slow increase due to the higher kinetic energy of generated hot electrons (also refereed as their "effective temperature", $T_h$). This is conditioned by a gentle monotonic increase of the Mo K-shell ionisation cross-section with the electron energy in the region of ≥100 keV. [24] Estimations show that the ponderomotive heating at laser intensity of $I_{las}$= $2\times10^{18}$ W/cm$^2$ results in the electron temperature of $T_h \approx$ 200 keV, according to a well-known scaling law. [11-13] On the other hand, more energetic electrons penetrate deeper in the target material, so that probability of a deeply-born Kα photon to escape from the target tends to fall down. Estimations show that electrons at energies of 0.1 and 1 MeV have penetration depth of approximately 20 and 400 μm, coordinately. [25] Whereas the absorption depth in molybdenum of a photon at 17.4 keV equals to ~50 μm. [26] So, it seems quite natural that the Kα conversion efficiency saturates at the relativistic intensities and even starts to decrease at $I_{las} \approx 1\times10^{19}$ W/cm$^2$.

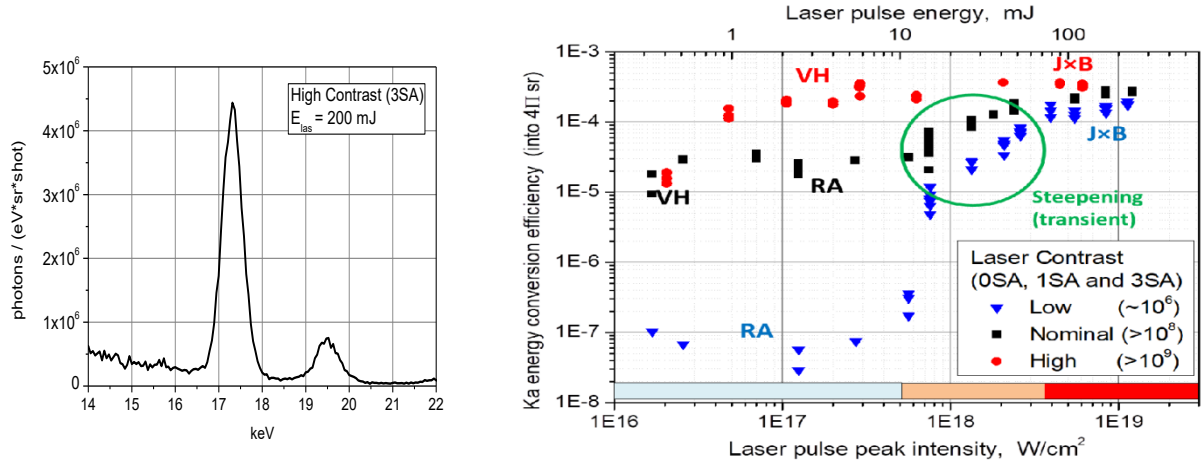

Fig. 3. Reconstructed absolute spectrum of the X-ray source Kα line emission. (to the left)
Fig. 4. Dependencies of the Kα conversion efficiency on the laser pulse peak intensity measured at different laser pulse temporal contrast. The marked intervals of laser intensity correspond to:    - non-relativistic,    - sub-relativistic, and    - relativistic laser intensity regimes.

### 3.1. Pre-formed plasma scale length

In fact, different behaviour of Kα conversion efficiency dependencies depicted in Fig.4 are signatures of the action of different physical mechanisms of the laser-plasma interaction. One of the key influencing parameters is the laser pulse contrast, which governs the formation of plasma at the target surface prior to the arrival of main high-intensity femtosecond pulse.

Let consider three different scenarios for the pre-plasma formation, which take place in our experiments and can be distinguished by comparison of the ASE pedestal intensity, $I_{ASE}$, with two characteristic values: (i) the ablation threshold of target material, $I_a \approx 5\times10^8$ W/cm$^2$, and (ii) the threshold of its direct ionization, $I_i \approx 10^{12}$ W/cm$^2$. [10,11,27]

These scenarios differ by the electron density scale length of the preformed plasma shell, $L$, at the level of critical electron density, $n_{ce}$, which are determined by the following expressions: $L = |\nabla n_e(z)/n_{ce}|^{-1}$, and $n_{ce} = \epsilon_0 m_e \omega_{las}^2 / e^2$, where $\epsilon_0$ is the vacuum permittivity and $z$ is coordinate along a normal to the target surface. The critical density corresponds to the condition, in which the own frequency of plasma density oscillations is equal to the laser frequency. The laser beam cannot propagate beyond the critical surface to the region of overdense plasma. On the other hand, the laser radiation pressure can violate the plasma density profile exactly in the region of critical density, where the laser beam is absorbed and reflected.

The first scenario is played when $I_{ASE} < I_a$. In this case, neither ablation, nor direct ionization of the target occurs, and the plasma shell is formed only by the directly ionizing high-intensity rising edge of the main fs pulse. This process acts on the ps time scale, so that heavy ions do not have enough time to move. Therefore, the preformed plasma density scale length may be estimated by the relation $L \ll \lambda$. In case of the high laser contrast (>10$^9$), this scenario takes place for the intensities of main fs laser pulse of <10$^{18}$ W/cm$^2$, while at the nominal contrast (>10$^8$), it acts for intensities restricted to <10$^{17}$ W/cm$^2$. It is not played at the low contrast in our experiments.

The second scenario is played, when $I_a < I_{ASE} < I_i$. In this case, the target is intensively ablated by the laser beam, forming the gaseous shell consisted of atoms and ions ejected from heated target surface. Although the vapour shell expands rather slowly, the process occurs on the ns time scale, which offers to ablated particles enough time to propagate over distances comparable with the laser wavelength. [28] However, this process cannot create a plasma with density exceeding the critical density of ~$1.7\times10^{21}$ cm$^{-3}$ corresponding to the Ti:Sa laser frequency. The dense electron preplasma is formed only when the plume particles are rapidly and multiply ionized (producing highly-stripped ions) by the intense optical field of the rising edge of main fs pulse. In this case, the plasma density scale length may reach values of $L\sim\lambda$. This scenario is played in part at all three contrast configurations and on the major interval of examined laser intensities.

According to the third (and the least) scenario, the target surface is directly ionised by the ASE pedestal, which reaches the intensity $I_i = 10^{12}$ W/cm$^2$ and higher. In this case, a hot ionized plume of pre-formed plasma is very rapidly expanding on the ns time scale, so that the relation $L>\lambda$ is valid. Concerning our experiments, it happens only in the case of low laser contrast and relativistic laser intensity.

**3.2. Acting interaction mechanism**

Now, let consider the second key influencing parameter: the laser pulse intensity. In fact, the influence of laser intensity can be formally separated to three strongly effecting factors: (i) the electron quiver amplitude, which determines whether electron can escape from plasma and be accelerated outside of it, (ii) the relativistic relation, $a_0>1$, which determines a strong influence of magnetic component of the Lorentz force on the electron motion and acceleration, and (iii) the laser radiation pressure, which can violate the plasma density profile at the critical surface.

Firstly, the above estimated scale length of the preformed plasma should be compared with the electron quiver amplitude in the electric field of the main femtosecond laser pulse. This amplitude is determined by the relation $A = eE_0/(m_e \omega^2_{las})$. Note that at $a_0=1$, one has $A = \lambda/2\pi$.

In the case of $A \leq L$, the electron cannot escape from the dense plasma. However, if the electric field component parallel to the direction of plasma density gradient is non-zero (as for obliquely incident P-polarised laser beam), the laser energy can be resonantly transmitted to the longitudinal plasma density wave, resulting in a high plasma wave amplitudes capable of accelerating electrons to energies of tens keV. Such kind of heating mechanism is called the resonance absorption (RA). [10-13] We attribute this mechanism to those parts of the K$\alpha$ efficiency dependencies in Fig.4, where the preformed plasma scale length can be significant (i.e. $L\sim\lambda$ or $L>\lambda$) and where the quiver amplitude is small ($A<<\lambda$). So, it takes place at the low and nominal contrast and the non-relativistic laser pulse intensities.

In the opposite case, $A > L$, the electron can be pulled out of plasma into the free space (or into the region of strongly rarefied plasma), where it can be accelerated back towards the target by the laser electric field (Brunel effect). [17] Such process is called the vacuum heating (VH). We can attribute this mechanism to those parts of the dependence for high contrast, for which we have estimated above that $L<<\lambda$, and where the non-relativistic and sub-relativistic intensity regimes take place. Note that a local increase of the K$\alpha$ conversion efficiency, which is observed at the nominal contrast in the region of low laser intensity of ~$3\times10^{16}$ W/cm$^2$, and its coincidence with the conversion efficiency values within this region at high contrast (see Fig.4), indicate to the switching from RA to VH mechanism. This idea is also supported by our above estimate of $L<<\lambda$ for the nominal contrast and non-relativistic laser intensity.

Secondly, as the electron quiver velocity approaches to the speed of light, the influence of magnetic component of the Lorentz force begins to manifest itself more and more in the electron motion. It guides to the dominant role of another plasma absorption mechanism called the ponderomotive or J×B heating. This mechanism is physically very similar to the VH, however, differently to it, is driven also by the oscillating component of the ponderomotive potential, which is responsible for $v\times B$ term of the Lorentz force. The resulting ponderomotive force oscillates twice the laser frequency and accelerates electrons along the axis of laser beam propagation. [10-13]

Thirdly, as approaching to the relativistic laser intensities, the laser radiation pressure takes the force. Indeed, it is proportional to the light intensity and may be expressed as $P \approx I/c$. So, at laser intensity of $3\times10^{18}$ W/cm$^2$, for example, the radiation pressure approaches to an enormously high value of 1 Gbar. It can drastically steepen the electron density profile at the region of critical density, thus favouring the transition from RA to VH and J×B mechanisms. The effect of radiation pressure on the preformed plasma is experimentally confirmed by the Doppler shift in the radiation wavelength of a probe laser beam, which is reflected by moving critical-density boundary. [29] Alternatively, it is observed by Doppler shift in the harmonics of the pump laser pulse, which are generated also at the moving critical surface. [30,31] The increasing role of the profile steepening is clearly distinguished in Fig.4 by the growth of K$\alpha$ conversion efficiency for the low and nominal pulse contrast within the interval of laser peak intensities of $5\times10^{17}$ - $5\times10^{18}$ W/cm$^2$, which corresponds to sub-relativistic regime and is marked by the brown band. It should be pointed out that similar dependence for the absorption coefficient of laser radiation by plasma has been observed experimentally and simulated numerically by other investigators. [31,32]

At the relativistic intensities, the laser beam is capable to crush the preformed plasma density profile at the level of critical density in case of $L\sim\lambda$ and $L\geq\lambda$, thus establishing the electron plasma density scale length of $L\ll\lambda$. According to estimations, it is quite possible that this process is accomplished on the femtosecond time scale. [10] Notably, Fig.4 illustrates the same consequences of strengthening of the radiation pressure for the sub-relativistic laser pulses, which differ by two orders in their temporal contrast. As a result, the J×B mechanism dominates other processes of electron heating at the relativistic interaction regime (marked by the red band) leading to almost equivalent Kα conversion efficiency for all three examined laser contrast values.

Finally, for more clarity, we have introduced in Fig.4 the labels with acronyms of the attributed laser-plasma absorption mechanism, which prevails the interaction in the corresponding conditions. On this basis we now can give a quantitative estimate of the pre-formed plasma density scale length $L$, comparing it with the actual value of quiver amplitude $A$, and applying the condition for functioning of the VH mechanism, which requires that $A>L$. Hence, at the laser pulse peak intensity of $I_{las}= 3\times10^{16}$ W/cm$^2$ (corresponds to $E_{las}= 0.6$ mJ on target) and the high and nominal contrast, we have $L< 0.014\times \lambda/2\pi \approx 2$ nm, while at $I_{las}= 5\times10^{17}$ W/cm$^2$ ($E_{las}= 10$ mJ) and the high contrast we have $L< 0.23\times \lambda/2\pi \approx 30$ nm. Whereas, at the nominal contrast and the latter intensity value, the inverse relation, $L> 0.23\times \lambda/2\pi \approx 30$ nm, should be valid. The effect of plasma profile steepening does not allow carrying out of analogous estimations to higher laser intensities. The hydrodynamic considerations are required in this case.

### 3.3. Dependence of Kα conversion efficiency on the laser beam incidence angle

Let remind that, in all above discussed experiments, the laser beam incidence angle to target is set to 45 degrees. Further, we verify the validity of the above attributed interaction mechanisms by examining dependencies of the Kα conversion efficiency on the angle $\theta$, which is determined between the normal to target surface and the laser beam propagation direction.

Experimental observations and simulations of the angular dependencies of laser energy absorption by plasma in different interaction regimes are discussed in many publications and are explicitly reviewed in Ref. [13] It is shown that, in the case of resonant excitation of plasma wave (RA) and validity of $L\sim\lambda$ or $L>\lambda$ conditions, the maximum values of plasma absorption correspond to $\theta$-angle lying in the interval of 20-30 degrees. Differently to that, the action of VH mechanism is characterised by a steep plasma density profile and the maximum absorption near 45-50 degrees. [33-35] Under the VH mechanism, the electric field drives acceleration of electrons in the transverse to laser beam propagation direction. Thus, the component of electron acceleration towards the target vanishes as approaching the normal laser incidence, resulting in a decrease in the Kα photon production. Oppositely, the oscillating ponderomotive force produced by the J×B mechanism accelerates electrons in the longitudinal direction, generating similar effects to those under VH mechanism, but at the doubled oscillation frequency. As a result, the electrons are accelerated in the both, longitudinal and transverse, directions during one laser cycle and follow trajectories with the form of number 8, if their drift velocity is zero. Effect of the ponderomotive force is equivalent to that produced by a driving electrostatic field with $E_{dr} = (m_e /4e) \cdot \partial(v_q^2(z))/\partial z \cdot \cos(2\omega_{las}t)$, where $v_q$ is the electron quiver velocity, while thermal and relativistic effects are neglected for simplicity. [36] Estimating the collisionless skin depth for the latter as $c/2\omega_{las}$, we obtain the relation: $E_{dr}/E_0 \approx a_0/2$. It shows that within the examined range of relativistic $a_0$ values varying from 1 to 5, the influence of both components of the Lorentz force is comparable. That should lead to almost independent Kα conversion efficiency on the laser incidence angle under action of J×B heating mechanism.

Fig.5 a), b) and c) demonstrate the observed angular dependencies of Kα conversion efficiency in the cases of high, nominal and low laser contrast, respectively. Figures a) and b) show experimental points measured at fixed laser pulse peak intensities of $1.2\times10^{17}$, $7.5\times10^{17}$, and $2.5\times10^{18}$ W/cm$^2$, which correspond to, respectively, non-relativistic ($a_0 = 0.05$), sub-relativistic ($a_0 = 0.3$), and relativistic ($a_0 = 1.1$) laser intensity regimes. In figures c) and d), the non-relativistic dependencies are replaced by those at the strongly-relativistic regime ($a_0 = 4.5$).

Note that Kα conversion efficiency measured at the angles approaching to normal and grazing incidence may significantly deviate from the real values due to a number of factors. Firstly, the laser beam is not collimated, but contained within the focusing cone with apex angle of 15 degrees. Secondly, the X-ray detection axis is oriented perpendicular to the laser axis (see Fig.2). So at small angles, it results in a partial shielding of the X rays, which are generated deeply in the target material. Thirdly, near the grazing incidence, the peak laser intensity rapidly degrades due to the defocusing on a large part of the focal spot imprinted on target surface, also, the component of radiation pressure normal to the plasma density gradient tends to zero, etc. Therefore, the interaction conditions no longer correspond to those assumed for given dependence at other incidence angle values.

Notably, all presented dependences show a remarkable accordance with the above discussed features inherent to the acting interaction mechanism that we have attributed in Fig.4 to different combinations of the laser pulse contrast and intensity. So, the two dependencies in Fig.5-a, which correspond to the non-relativistic and sub-relativistic regimes and high contrast, confirm operation of the VH mechanism in these conditions by exhibiting maximum efficiency near incidence angle of 45°. Whereas the third dependence shown in this figure, which is obtained at the relativistic laser intensity, appears to be almost independent on the angle, thus indicating the action of J×B heating mechanism.

Angular dependencies at the nominal contrast (Fig.5-b) demonstrate that the maximum of Kα photon production at the non- and sub-relativistic laser intensities is shifted to $\theta \approx 25°$, as it is inherent for the RA mechanism. It is not surprising, because of the increased plasma density scale length $L$ due to an order lower pulse contrast with respect to that for Fig. 5-a. Note that a significant positive variation is seen for the considered dependences near the angle of $\theta=55°$, i. e. close to the maximum absorption under VH mechanism. The general form of these dependences points to the switching of dominant operating mechanism from RA to VH somewhere close to 45°. It confirms our conclusion above on the origin of the local increase in Kα efficiency at the nominal contrast and laser intensity of $\sim 3 \times 10^{16}$ W/cm$^2$ (see Fig.4). Also, a comparison of these two dependencies indicates that the relation between $A$ and $L$, taken at a fixed $\theta$ value, does not vary drastically for the two considered intensity regimes, thus reflecting the fact that the both parameters, $A$ and $L$, increase with the laser intensity. However, the effect of plasma profile steepening clearly manifests itself at the sub-relativistic regime, decreasing $L$ and strongly favoring the VH mechanism, especially at large $\theta$ values. The angular dependence in Fig.5-b, which is observed at the relativistic laser intensity, seems to be practically the same as that in Fig.5-a, also indicating to the operation of J×B mechanism in this regime.

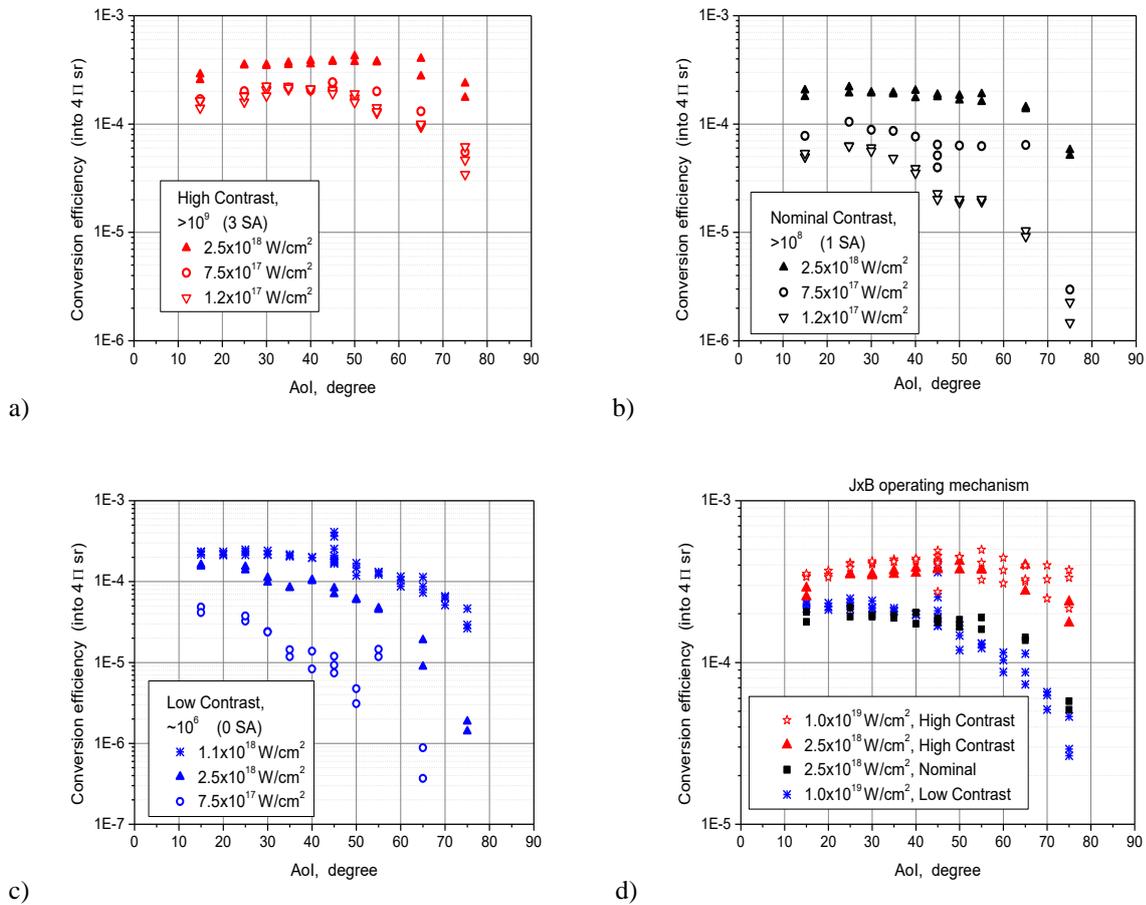

Fig.5. Dependencies of the Kα conversion efficiency on the angle of incidence of laser beam to target: a) at high laser contrast; b) at nominal laser contrast; c) at low laser contrast; and d) for J×B operating mechanism.

The results obtained at low contrast (~$10^6$) are shown in Fig.5-c. The corresponding values of Kα conversion efficiency at non-relativistic regime are generally lower than $10^{-7}$ and are not seen in the figure. Instead, the dependence obtained at strongly-relativistic laser intensity ($a_0 = 4.5$) is introduced. It is seen that dependencies related to the sub-relativistic and relativistic regimes point to the action of the RA mechanism, while that observed for the strongly relativistic regime seems to indicate to the J×B heating. Naturally, the latter dependence resembles that shown in Fig.5-b for the nominal contrast and obtained at relativistic intensity regime.

For the convenience of comparison, the dependencies observed at relativistic and strongly relativistic regimes are reproduced in a separate figure. It is seen in Fig.5-d that dependencies correspondent to the high contrast are very similar to each other and almost independent on the angle of incidence of laser radiation. However, two other dependences presented in the figure, being also similar to each other, differ from the two previous by a significant decrease in the efficiency at large angles. This effect can be explained by important losses of the laser radiation due to its scattering and partial absorption along longer distances into a far-extending layer of rarified plasma. In fact, even being strongly steepened at the critical surface, the density profile of a thick pre-formed plasma exhibits a shallow but far extending tail. [12] Also, higher reflection at grazing incidence can be considered. Taking into account these factors, we confirm the attribution of all dependences introduced in Fig.5-d to the J×B heating mechanism.

### 3.4. X-ray source size measurements

The technique of knife-edge shadowgraphy is applied to measure the X-ray source size. [16] The obtained results are presented in Fig.6, demonstrating strong dependence on the laser pulse contrast and intensity.

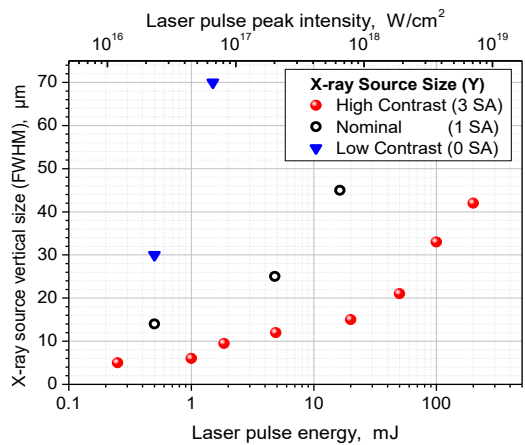

Fig.6. Dependence of the X-ray source size on the laser pulse peak intensity and contrast.

The source size of ~10 μm is observed at the high contrast values of >$10^9$. As seen in the figure, lower values of the laser pulse temporal contrast result in much larger X-ray emitting spot size, thus restricting many important applications.

## CONCLUSION

This paper briefly describes the design and main technical parameters of the intense laser-produced X-ray source of molybdenum Kα line emission at 17.4 keV, as well as the applied X-ray diagnostics. High values of the Kα conversion efficiency of up to $2\times10^{-4}$ into 2π sr are obtained, which correspond to the best results reported to date.

The main emphasis in the work is made to systematic experimental studies of the conversion efficiency of the laser radiation to the Kα line X-ray emission, which are performed within the wide ranges of the laser pulse peak intensities and temporal contrast. The key role of these two laser parameters on the physics of laser-plasma interaction is explicitly revealed. The effect of laser radiation pressure on the pre-formed plasma density profile is also clearly demonstrated. The performed simple and intuitive (but rather comprehensive) physical analysis allows unambiguous attribution of the Kα conversion efficiency dependencies observed at different combinations of the key laser parameters to one of three main interaction mechanisms. Our results confirm a general notion that J×B heating must dominate the absorption physics at relativistic laser intensities. [13] The X-ray source generated by high-intensity fs pulses with the high temporal contrast of >$10^9$ demonstrate a rather small size of few tens µm, which is favourable for many important applications.

We consider this publication as an important step in the diffusion of our new experimental results. Certainly, a more accurate analysis, including more precise estimations supported by physical modelling and numerical simulations, will be provided in near future. Also, we plan to discuss technical aspects of the applied experimental methods in more details in a separate publication.

## ACKNOWLEDGMENTS

Financial support of the European Community, Ministry of Research and High Education, Region Provence-Alpes-Côte d'Azur, Department of Bouches-du-Rhône, City of Marseille, CNRS, and Aix-Marseille University is gratefully acknowledged for funding ASUR platform. We thank J.-C. Kieffer and S. Fourmeaux of INRS (Varennes, Canada) for fruitful discussions during the starting phase of this research program. We are also very thankful to R. Clady for the conditioning of the ASUR laser beam for our work and for his help, with A. Ferre, in the preparation and carrying out the experiments, to L. Charmasson and O. Uteza for their help in the designing of the experimental setup and implementation of the radioprotection control during the works. Our particular acknowledgements are addressed to M. Sentis who has conceived and supervised the works on the development of Mo Kα-line X-ray source at the LP3 Laboratory.